\documentstyle[12pt]{article} 

\makeatletter
\@addtoreset{equation}{section}
\makeatother
\def\theequation{\thesection.\arabic{equation}}
\def\thefootnote{%
\fnsymbol{footnote}}

\newcommand{\dps}{\displaystyle }
\newcommand{\e }{\varepsilon }
\newcommand{\et}{\tilde{\varepsilon }}
\newcommand{\al }{\alpha }
\newcommand{\de }{\delta }

\newcommand{\ket }{\rangle }
\newcommand{\bra }{\langle }
\newcommand{\ga }{\gamma }
\newcommand{\la }{\lambda }
\newcommand{\Hw }{H_{\hbox{w}}} 
\newcommand{\im}{\hbox{Im}}
\newcommand{\re}{\hbox{Re}}
\newcommand{\La }{\Lambda }
\newcommand{\De }{\Delta }
\newcommand{\no }{\nonumber }
\newcommand{\kob }{\overline{K^0}}
\newcommand{\ko }{K^0}
\newcommand{\nub }{\overline{\nu }}
\newcommand{\fb }{\overline{f}}
\newcommand{\xib }{\overline{\xi }}

\begin{document}
\begin{flushright}
NUP-A-98-10\\April 1998
\end{flushright}
~\\ ~\\ 
\begin{center}
\Large{Further Study on Possible Violation of $CP$, $T$ and $CPT$ Symmetries in the $\ko$-$\kob$ System \\ 
--- Remarks and Results ---\footnote{Revised version of the report presented by S.Y.Tsai at the Workshop on Fermion Masses and CP Violation (Hiroshima, Japan, March 5-6, 1998).} } 
\end{center} 
~\\ ~\\ 
\begin{center}
Yutaka Kouchi, Akihiro Shinbori, Yoshihiro Takeuchi and S. Y. Tsai\footnote{E-mail address: tsai@phys.cst.nihon-u.ac.jp}
\end{center}

\begin{center}
{\it Atomic Energy Research Institute and Department of Physics \\ 
College of Science and Technology, Nihon University \\ 
Kanda-Surugadai, Chiyoda-ku, Tokyo 101, Japan }
\end{center}
~\\ ~\\ 


\begin{abstract}

We demonstrate how one may identify or constrain possible violation of $CP$, $T$ and $CPT$ symmetries in the $\ko $-$\kob $ system in a way as phenomenological and comprehensive as possibie. For this purpose, we first introduce parameters which represent violation of these symmetries in mixing parameters and decay amplitudes in a well-defined way. After discussing some characteristics of these parameters, we derive formulae which relate them to the experimentally measured quantities. We then carry out a numerical analysis with the help of the Bell-Steinberger relation to derive constraints to these violating parameters from available experimental data. Finally, we compare our parametrization and procedure of analysis with those employed in the recent literature.
\end{abstract}

\newpage 

\setcounter{footnote}{0}
\def\thefootnote{%
\mbox{\alph{footnote}}}

\section{Introduction}

~~~~Although, on the one hand, the standard field theory implies that $CPT$ symmetry should hold exactly, and, on the other hand, all experimental observations up to now are perfectly consistent with this symmetry, continued experimental, phenomenological and theoretical studies of this and related symmetries are warrented.
 
In a series of  papers[1-5], we have demonstrated how one may identify or constrain possible violation of $CP$, $T$ and $CPT$ symmetries in the $\ko $-$\kob $ system in a way as phenomenological and comprehensive as possible. For this purpose, we have first introduced parameters which represent violation of these symmetries in mixing parameters and decay amplitudes in a well-defined way and related them to the experimentally measured quantities. We have then carried out a numerical analysis with little theoretical input to derive constraints to these violating parameters from available experimental data. It has been shown among other things that the most recent results on leptonic asymmetries obtained by the CPLEAR Collaboration[6]  allow one for the first time to constrain to some extent possible $CPT$ violation in leptonic decay modes.\footnote{After our last paper[5] being sent to the major high energy physics centers, we became aware that the CPLEAR Collaboration themselves[7] had also, by an analysis more or less similar to ours, reached the similar conclusion.}

As discussed in [1-3], our parametrization is very unique in that it is manifestly invariant with respect to rephasing of the $|\ko \ket $ and $|\kob \ket $ states, 
\begin{equation}
|\ko \ket \to |\ko \ket '=|\ko \ket e^{-i\xi _K}~,\qquad |\kob \ket \to |\kob \ket '=|\kob \ket e^{i\xi _K}~.
\end{equation}
It has to be noted however that our parametrization is not invariant with respect to rephasing of final states $|f\ket $, e.g. 
\begin{equation}
\left. 
	\begin{array}{ccc}
	|(2\pi )_I\ket &\to &|(2\pi )_I\ket '=|(2\pi )_I\ket e^{-i\xi _I}~,\\ \\
	|\ell ^-\ket &\to &|\ell ^-\ket '=|\ell ^-\ket e^{-i\xi _{\ell }}~,\\ \\
	|\ell ^+\ket &\to &|\ell ^+\ket '=|\ell ^+\ket e^{-i\xib _{\ell }}~.
	\end{array}
\right. 
\end{equation}
where $I=0$ or $2$ stands for the isospin of the $2\pi $ states, $|\ell ^-\ket =|\pi ^+\ell ^-\nub_{\ell }\ket $, $|\ell ^+\ket =|\pi ^-\ell ^+\nu_{\ell }\ket $ and $\ell =e$ or $\mu$. We have rather adopted a specific phase convention for the final states $|f\ket $. Some of the constraints which we have claimed to follow from $CP$, $T$ and/or $CPT$ symmetries do depend on this phase convention and have to be distinguished from those constraints which are phase-convention-independent. We would like to clarify these points in Sec.~4 of the present work.

The main data we have used in our last analysis[5] are those reported by the CPLEAR Collaboration [6] and those compiled by the Particle Data Group in [8]. The latter article also contains a number of notes giving definition of parameters, formulae relevant for data processing and related remarks. In Sec.~7, we would like to compare our parametrization and procedure of analysis with those given or cited in [8].

To be self-contained, we need to recapitulate our parametrization(Sec.~2 and Sec.~3), formulae(Sec.~5) and main results(Sec.~6). These parts are essentially same with our previous works[1-5], except that the $\pi ^+\pi ^-\ga $ state is taken into account as one of intermediate states in the Bell-Steinberger relation.


\section{The $\ko$-$\kob $ mixing and the Bell-Steinberger relation}

~~~~Let $|\ko \ket $ and $|\kob \ket $ be eigenstates of the strong interaction with strangeness $S=+1$ and $-1$, related to each other by $(CP)$, $(CPT)$ and $T$ operations as[1, 2, 9]
\begin{equation}
\left. 
	\begin{array}{cc}
	(CP)|\ko \ket =e^{i\al _K}|\kob \ket ~,& (CPT)|\ko \ket =e^{i\beta _K}|\kob \ket ~,\\ \\ 
	(CP)|\kob \ket =e^{-i\al _K}|\ko \ket ~,& (CPT)|\kob \ket =e^{i\beta _K}|\ko \ket ~,\\ \\ 
	T|\ko \ket =e^{i(\beta _K-\al _K)}|\ko \ket ~,&T|\kob \ket =e^{i(\beta _K+\al _K)}|\kob \ket ~.
	\end{array}
\right. 
\end{equation}
Note here that, given the first two where $\al _K$ and $\beta _K$ are arbitrary real parameters, the rest follow from $(CP)T=T(CP)=(CPT)$ and anti-linearity of $T$ and $(CPT)$.

When the weak interaction $\Hw $ is switched on, the $\ko $ and $\kob $ states decay into other states and become mixed. The time evolution of the arbitrary state 
\[
|\Psi (t)\ket =c_1(t)|\ko \ket +c_2(t)|\kob \ket 
\]
is described by a Schr\"odinger-like equation[10] 
\begin{equation}
i\frac{d}{dt}
\left( 
	\begin{array}{c}
	c_1(t)\\ \\ c_2(t) 
	\end{array}
\right) 
=\La 
\left( 
	\begin{array}{c}
	c_1(t)\\ \\ c_2(t) 
	\end{array}
\right) ~. 
\end{equation}
$\La =(\La _{ij})$ is a $2\times 2$ matrix related to $\Hw $, e.g. 
\[
\La _{12}=\sum _f\bra \ko |\Hw |f\ket \bra f|\Hw |\kob \ket /(m_K-E_f+i\e )~,
\]
and may be written as 
\begin{equation}
\La =M-i\frac{\Gamma }{2}~,
\end{equation}
$M(\Gamma )$ being an hermitian matrix called mass (decay) matrix. The two eigenstates of $\La $ and their respective eigenvalues may be written as 
{
\setcounter{enumi}{\value{equation}}
\addtocounter{enumi}{1}
\setcounter{equation}{0}
\renewcommand{\theequation}{\thesection.\theenumi\alph{equation}}
\begin{eqnarray}
|K_S\ket &=&\frac{1}{\sqrt{|p_S|^2+|q_S|^2}}\left( p_S|\ko \ket +q_S|\kob \ket \right) ~, 
\end{eqnarray}
\begin{eqnarray}
|K_L\ket &=&\frac{1}{\sqrt{|p_L|^2+|q_L|^2}}\left( p_L|\ko \ket -q_L|\kob \ket \right) ~; 
\end{eqnarray}
\setcounter{equation}{\value{enumi}}}%
{
\setcounter{enumi}{\value{equation}}
\addtocounter{enumi}{1}
\setcounter{equation}{0}
\renewcommand{\theequation}{\thesection.\theenumi\alph{equation}}
\begin{equation}
\la _S=m_S-i\frac{\ga _S}{2}~,
\end{equation}
\begin{equation}
\la _L=m_L-i\frac{\ga _L}{2}~.
\end{equation}
\setcounter{equation}{\value{enumi}}}%
$\la _S$, $\la _L$, $q_S/p_S$ and $q_L/p_L$ are related to $\La _{ij}$, and $m_{S, L}=\re (\la _{S, L})$ and $\ga _{S, L}=-2\im (\la _{S, L})$ are the mass and the total decay width of the $K_{S, L}$ state respectively. Parametrizing $q_S/p_S$ and $q_L/p_L$ as 
\begin{equation}
\left. 
	\begin{array}{c}
	\frac{\dps{q_S}}{\dps{p_S}}=e^{i\al _K}\frac{\dps{1-\e _S}}{\dps{1+\e _S}}~,\\ \\ 
	\frac{\dps{q_L}}{\dps{p_L}}=e^{i\al _K}\frac{\dps{1-\e _L}}{\dps{1+\e _L}}~,
	\end{array}
\right. 
\end{equation}
one may express $|K_S\ket $ and $|K_L\ket $ as 
{
\setcounter{enumi}{\value{equation}}
\addtocounter{enumi}{1}
\setcounter{equation}{0}
\renewcommand{\theequation}{\thesection.\theenumi\alph{equation}}
\begin{eqnarray}
|K_S\ket &=&\frac{1}{\sqrt{2(1+|\e _S|^2)}}\left\{ (1+\e _S)e^{-i\al _K/2}|\ko \ket +(1-\e _S)e^{i\al _K/2}|\kob \ket \right\} \no \\ 
&=&\frac{1}{\sqrt{1+|\e _S|^2}}\left( |K_1\ket +\e _S|K_2\ket \right)~,
\end{eqnarray}
\begin{eqnarray}
|K_L\ket &=&\frac{1}{\sqrt{2(1+|\e _L|^2)}}\left\{ (1+\e _L)e^{-i\al _K/2}|\ko \ket -(1-\e _L)e^{i\al _K/2}|\kob \ket \right\} \no \\ 
&=&\frac{1}{\sqrt{1+|\e _L|^2}}\left( |K_2\ket +\e _L|K_1\ket \right)~,
\end{eqnarray}
\setcounter{equation}{\value{enumi}}}%
where 
\begin{equation}
|K_{1,2}\ket =\frac{1}{\sqrt{2}}\left( e^{-i\al _K/2}|\ko \ket \pm e^{i\al _K/2}|\kob \ket \right) 
\end{equation}
are $CP$ eigenstates with $CP=\pm 1$. Note that the overall phases of $|K_{1,2}\ket $ and $|K_{S,L}\ket $ are chosen in such a way as[1]
\[
CPT|K_{1,2}\ket =\pm e^{i\beta _K}|K_{1,2}\ket ~, 
\]
\[
|K_{S,L}\ket \to |K_{1,2}\ket \quad \mbox{as} \quad \e _{S,L} \to 0~.
\]
$\e _{S,L}$ will further be parametrized as 
\begin{equation}
\e _{S,L}=\e \pm \de ~.
\end{equation}

From the eigenvalue equation of $\La $, one may readily derive the well-known Bell-Steinberger relation[11]:
\begin{equation}
\left[ \frac{\ga _S+\ga _L}{2}-i\De \right] \bra K_S|K_L\ket = \bra K_S|\Gamma |K_L\ket ~,
\end{equation}
where 
\begin{equation}
\bra K_S|\Gamma |K_L\ket = 2\pi \sum _f\bra K_S|\Hw|f\ket \bra f|\Hw |K_L\ket \de (m_K-E_f)~,
\end{equation}
\begin{equation}
\De =m_S-m_L~.
\end{equation}
One may further verify[3, 4]
{
\setcounter{enumi}{\value{equation}}
\addtocounter{enumi}{1}
\setcounter{equation}{0}
\renewcommand{\theequation}{\thesection.\theenumi\alph{equation}}
\begin{equation}
\e _{\| }\equiv \re [\e \exp (-i\phi _{SW})]\simeq \frac{-2\im (M_{12}e^{i\al _K})}{\sqrt{(\ga _S-\ga _L)^2+4\De ^2}}~,
\end{equation}
\begin{equation}
\e _{\perp }\equiv \im [\e \exp (-i\phi _{SW})]\simeq \frac{\im (\Gamma _{12}e^{i\al _K})}{\sqrt{(\ga _S-\ga _L)^2+4\De ^2}}~,
\end{equation}
\setcounter{equation}{\value{enumi}}}%
{
\setcounter{enumi}{\value{equation}}
\addtocounter{enumi}{1}
\setcounter{equation}{0}
\renewcommand{\theequation}{\thesection.\theenumi\alph{equation}}
\begin{equation}
\de _{\|}\equiv \re [\de \exp (-i\phi _{SW})]\simeq \frac{(\Gamma _{11}-\Gamma _{22})}{2\sqrt{(\ga _S-\ga _L)^2+4\De ^2}}~,
\end{equation}
\begin{equation}
\de {\perp }\equiv \im [\de \exp (-i\phi _{SW})]\simeq \frac{(M_{11}-M_{22})}{\sqrt{(\ga _S-\ga _L)^2+4\De ^2}}~,
\end{equation}
\setcounter{equation}{\value{enumi}}}%
where 
\begin{equation}
\phi _{SW}=\tan ^{-1}\left( \frac{-2\De }{\ga _S-\ga _L} \right) 
\end{equation}
is the so-called superweak phase. 


\section{Decay amplitudes }

~~~~The $\ko $ and $\kob $ (or $K_S$ and $K_L$) states have many decay channels, among which we concentrate on the following four relevant modes. 


\subsection{$2\pi $ modes }

~~~~We parametrize amplitudes for $\ko $ and $\kob $ to decay into $(2\pi )_I$ as[1]
\begin{equation}
\left. 
	\begin{array}{c}
	\bra (2\pi )_I|\Hw |\ko \ket =F_I(1+y_I)e^{i\al _K/2}~, \\ \\ 
	\bra (2\pi )_I|\Hw |\kob \ket =F_I^*(1-y_I^*)e^{-i\al _K/2}~,
	\end{array}
\right. 
\end{equation}
and further introduce 
\begin{equation}
z_I=\frac{\im (F_I)}{\re (F_I)}~. 
\end{equation}

The experimentally measured quantities are $\eta _{+-}$ and $\eta _{00}$ defined by 
{
\setcounter{enumi}{\value{equation}}
\addtocounter{enumi}{1}
\setcounter{equation}{0}
\renewcommand{\theequation}{\thesection.\theenumi\alph{equation}}
\begin{equation}
\eta _{+-}=|\eta _{+-}|e^{i\phi _{+-}}=\frac{\bra \pi ^+\pi ^-,\mbox{outgoing}|\Hw |K_L\ket }{\bra \pi ^+\pi ^-,\mbox{outgoing}|\Hw |K_S\ket }~,
\end{equation}
\begin{equation}
\eta _{00}=|\eta _{00}|e^{i\phi _{00}}=\frac{\bra \pi ^0\pi ^0,\mbox{outgoing}|\Hw |K_L\ket }{\bra \pi ^0\pi ^0,\mbox{outgoing}|\Hw |K_S\ket }~.
\end{equation}
\setcounter{equation}{\value{enumi}}}%
Defining 
\begin{equation}
\eta _I=\frac{\bra (2\pi )_I|\Hw |K_L\ket }{\bra (2\pi )_I|\Hw |K_S\ket }~,
\end{equation}
\begin{equation}
\omega =\frac{\bra (2\pi )_2|\Hw |K_S\ket }{\bra (2\pi )_0|\Hw |K_S\ket }~,
\end{equation}
one gets 
{
\setcounter{enumi}{\value{equation}}
\addtocounter{enumi}{1}
\setcounter{equation}{0}
\renewcommand{\theequation}{\thesection.\theenumi\alph{equation}}
\begin{equation}
\eta _{+-}=\frac{\eta _0+\eta_2\omega '}{1+\omega '}~,
\end{equation}
\begin{equation}
\eta _{00}=\frac{\eta _0-2\eta _2\omega '}{1-2\omega '}~,
\end{equation}
\setcounter{equation}{\value{enumi}}}%
where 
\begin{equation}
\omega '=\frac{1}{\sqrt{2}}\omega e^{i(\de _2-\de _0)}~,
\end{equation}
$\de _I$ being the S-wave $\pi \pi $ scattering phase shift for the isospin $I$ state at an energy of the rest mass of $\ko $. $\omega $ is a measure of deviation from the $\De I=1/2$ rule, and may be inferred, for example, from 
\begin{eqnarray}
r&\equiv &\frac{\ga _S(\pi ^+\pi ^-)-2\ga _S(\pi ^0\pi ^0)}{\ga _S(\pi ^+\pi ^-)+\ga _S(\pi ^0\pi ^0)} \no \\ 
&=& \frac{4\re (\omega ')-2|\omega '|^2}{1+2|\omega '|^2} ~.
\end{eqnarray}
Here and in the following, $\ga _{S,L}(f)$ denotes the partial width for $K_{S,L}$ to decay into the final state $f$.


\subsection{$3\pi $ and $\pi ^+\pi ^-\ga $ modes }

~~~~The experimentally measured quantities are 
{
\setcounter{enumi}{\value{equation}}
\addtocounter{enumi}{1}
\setcounter{equation}{0}
\renewcommand{\theequation}{\thesection.\theenumi\alph{equation}}
\begin{equation}
\eta _{+-0}= \frac{\bra \pi ^+\pi ^-\pi ^0,\mbox{outgoing}|\Hw |K_S\ket }{\bra \pi ^+\pi ^-\pi ^0,\mbox{outgoing}|\Hw |K_L\ket }~,
\end{equation}
\begin{equation}
\eta _{000}= \frac{\bra \pi ^0\pi ^0\pi ^0,\mbox{outgoing}|\Hw |K_S\ket }{\bra \pi ^0\pi ^0\pi ^0,\mbox{outgoing}|\Hw |K_L\ket }~,
\end{equation}
\setcounter{equation}{\value{enumi}}}%

\begin{equation}
\eta _{+-\ga }= \frac{\bra \pi ^+\pi ^-\ga ,\mbox{outgoing}|\Hw |K_L\ket }{\bra \pi ^+\pi ^-\ga ,\mbox{outgoing}|\Hw |K_S\ket }~.
\end{equation}

We shall treat the $3\pi $ $(\pi ^+\pi ^-\ga )$ states as purely $CP$-odd ($CP$-even). 


\subsection{Leptonic modes }

~~~~We parametrize amplitudes for $\ko $ and $\kob $ to decay into $|\ell ^+\ket =|\pi ^-\ell ^+ \nu _{\ell }\ket $ and $|\ell ^-\ket =|\pi ^+\ell ^- \nub _{\ell }\ket $, where $\ell =e$ or $\mu $, as[1, 12, 13]
\begin{equation}
\left. 
	\begin{array}{l}
	\bra \ell ^+|\Hw |\ko \ket =F_{\ell }(1+y_{\ell })e^{i\al _K/2}~,\\ \\ 
	\bra \ell ^-|\Hw |\kob \ket =F_{\ell }^*(1-y_{\ell }^*)e^{-i\al _K/2}~, \\ \\ 
	\bra \ell ^+|\Hw |\kob \ket =x_{\ell +}F_{\ell }(1+y_{\ell })e^{-i\al _K/2}~, \\ \\ 
	\bra \ell ^-|\Hw |\ko \ket =x_{\ell -}^*F_{\ell }^*(1-y_{\ell }^*)e^{i\al _K/2}~.
	\end{array}
\right. 
\end{equation}
$x_{\ell \pm }$, which measure deviation from the $\De S=\De Q$ rule, are further parametrized as 
\begin{equation}
x_{\ell \pm }=x_{\ell }\pm x'_{\ell }~.
\end{equation}

Rather than the well measured time-independent asymmetry parameter 
{
\setcounter{enumi}{\value{equation}}
\addtocounter{enumi}{1}
\setcounter{equation}{0}
\renewcommand{\theequation}{\thesection.\theenumi\alph{equation}}
\begin{equation}
d^{\ell }_L = \frac{\ga _L(\pi ^-\ell ^+\nu _{\ell })-\ga _L(\pi ^+\ell ^-\nub _{\ell })}{\ga _L(\pi ^-\ell ^+\nu _{\ell })+\ga _L(\pi ^+\ell ^-\nub _{\ell })}~,
\end{equation}
the CPLEAR Collaboration[6] have for the first time measured two kinds of time-dependent asymmetry parameters 
\begin{equation}
d^{\ell }_1(t)=\frac{|\bra \ell ^-|\Hw |\ko (t)\ket |^2-|\bra \ell ^+|\Hw |\kob (t)\ket |^2}{|\bra \ell ^-|\Hw |\ko (t)\ket |^2+|\bra \ell ^+|\Hw |\kob (t)\ket |^2}~, 
\end{equation}
\begin{equation}
d^{\ell }_2(t)=\frac{|\bra \ell ^+|\Hw |\ko (t)\ket |^2-|\bra \ell ^-|\Hw |\kob (t)\ket |^2}{|\bra \ell ^+|\Hw |\ko (t)\ket |^2+|\bra \ell ^-|\Hw |\kob (t)\ket |^2}~.
\end{equation}
\setcounter{equation}{\value{enumi}}}%


\section{Conditions imposed by $CP$, $T$ and/or $CPT$ symmetries }

~~~~Although our amplitude parameters $F_f$ and $y_f$ as well as our mixing parameters $\e $ and $\de $ are all invariant with respect to rephasing of the $|\ko \ket $ and $|\kob \ket $ states, Eq.(1.1), $F_f$ and $y_f$ are not invariant with respect to rephasing of the final state $|f\ket $, Eq.(1.2). So are the relative $CP$ phase $\al _{\ell }$ between $|\ell ^+\ket $ and $|\ell ^-\ket $ and the relative $CPT$ phase $\beta _f$ between $|f\ket $ and $|\fb \ket $ defined in such a way as 
\begin{equation}
CP|\ell ^+\ket = e^{i\al _{\ell }}|\ell ^-\ket ~,\qquad CPT|f\ket =e^{i\beta _f}|\fb \ket ~,
\end{equation}
where $\al _{\ell }$ and $\beta _f$ are arbitrary real parameters and it is understood that $|\fb \ket =|f\ket $ for $|f\ket =|(2\pi )_I\ket $ and $|\pi ^+\pi ^- \ga \ket $ and $|\fb \ket =-|f\ket $ for $|f\ket =|\pi ^+\pi ^-\pi ^0\ket $ and $|\pi ^0 \pi ^0 \pi ^0\ket $. One may verify that $CP$, $T$ and $CPT$ symmetries impose such conditions as\footnote{In our previous papers[2-4], having adopted from the outset the phase convention 
{
\renewcommand{\theequation}{\thesection.6}
\begin{equation}
\al _{\ell }=0~,\qquad \beta _f=\beta _K~,
\end{equation}
\setcounter{equation}{\value{enumi}}}%
\setcounter{equation}{1}
we are led to claim that Eq.(4.4), too, would follow from $CP$, $T$ and/or $CPT$ symmetries and hence include $\im (F_{\ell })$ and $\im (y_f)$ in our list of symmetry-violating parameters. Also, our classification of $\e $ and $\de $ as indirect parameters and of $z_I$, $\re (y_f)$, $\im (x_{\ell })$ and $x'_{\ell }$ (and, erroneously, $\im (F_{\ell })$ and $\im (y_f)$ as well) as direct parameters is not very consistent, since non-vanishing of the latter set of parameters will in general result in non-vanishing of the former set of parameters. More consistent is to refer to $\e _{\| }$ and $\de _{\perp }$, which are related exclusively to the mass matrix, as indirect parameters and all the others, which are related to decay amplitudes or decay matrix, as direct parameters. 
}
{
\setcounter{enumi}{\value{equation}}
\addtocounter{enumi}{1}
\setcounter{equation}{0}
\renewcommand{\theequation}{\thesection.\theenumi\alph{equation}}
\begin{eqnarray}
	CP~\mbox{symmetry} &:& \im (F_{\ell })/\re (F_{\ell }) = -\tan \al _{\ell }/2~; \\ \no \\ 
	T~\mbox{symmetry} &:& 2\im (y_f)/(1-|y_f|^2)=\tan (\beta _f-\beta _K)~, \\ \no \\ 
	&& \im [(1+iz_I)^2(1-y_I^2)]=0~,\\ \no \\ 
	&& \im [F_{\ell }^2(1-y_{\ell }^2)\exp (i\al _{\ell })]=0~; \\ \no \\ 
	CPT~\mbox{symmetry} &:& \im (y_f)=\tan (\beta _f-\beta _K)/2~, 
\end{eqnarray}
\setcounter{equation}{\value{enumi}}}%
in addition to 
\begin{equation}
\left. 
	\begin{array}{ccl}
	CP~\mbox{symmetry} &:& \e =0,~\de =0,~z_I=0,~\re (y_f)=0,~\im (x_{\ell })=0,~\re (x'_{\ell })=0~; \\ \\ 
	T~\mbox{symmetry} &:& \e =0,~\im (x_{\ell })=0,~\im (x'_{\ell })=0 ~; \\ \\ 
	CPT~\mbox{symmetry} &:& \de =0,~\re (y_f)=0,~\re (x'_{\ell })=0,~\im (x'_{\ell })=0 ~. 
	\end{array}
\right. 
\end{equation}
One sees from Eqs.(4.2a,b,d,e) that, since $\al_{\ell }$ and $\beta _f$ are completely arbitrary, $\im (F_{\ell })$ and $\im (y_f)$ remain unconstrained even if one impose $CP$, $T$ and/or $CPT$ symmetries. It can be shown[14] however that it is possible by a choice of phase convention to set\footnote{See Ref.[12, 13] for related discussion on (non)observability of $\im (y_{\ell })$. } 
\begin{equation}
\im (F_{\ell })=0~,\qquad \im (y_f)=0~. 
\end{equation}
Eq.(4.2c) then gives 
\begin{equation}
T~\mbox{symmetry} : z_I=0~. 
\end{equation}


\section{Formulae relevant for numerical analysis }

~~~~We shall adopt a phase convention which gives Eq.(4.4). Observed and expected smallness of violation of $CP$, $T$ and $CPT$ symmetries and of the $\De S=\De Q$ rule allows us to treat all our symmetry-violating parameters (i.e. those implicit in Eqs.(4.3) and (4.5)), and $\re (x_{\ell })$ as well, as small, \footnote{As a matter of fact, we have already assumed that $CP$, $T$ and $CPT$ violations are small in deriving Eqs.(2.13a) $ \sim $ (2.14b). } and, to the leading order, one finds 
\begin{equation}
\omega \simeq \frac{\re (F_2)}{\re (F_0)}~,
\end{equation}
{
\setcounter{enumi}{\value{equation}}
\addtocounter{enumi}{1}
\setcounter{equation}{0}
\renewcommand{\theequation}{\thesection.\theenumi\alph{equation}}
\begin{equation}
\eta _I\simeq \e -\de +\re (y_I)+iz_I~,
\end{equation}
\begin{equation}
\eta _2-\eta _0\simeq \re (y_2-y_0)+i(z_2-z_0)~,
\end{equation}
\setcounter{equation}{\value{enumi}}}%
{
\setcounter{enumi}{\value{equation}}
\addtocounter{enumi}{1}
\setcounter{equation}{0}
\renewcommand{\theequation}{\thesection.\theenumi\alph{equation}}
\begin{equation}
d_1^{\ell }(t\gg 1/\ga _S)\simeq -4\re (\e )-2\re (y_{\ell }-x'_{\ell })~,
\end{equation}
\begin{equation}
d_2^{\ell }(t\gg 1/\ga _S)\simeq -4\re (\de )+2\re (y_{\ell }-x'_{\ell })~.
\end{equation}
\setcounter{equation}{\value{enumi}}}%
Furthermore, by taking $2\pi $, $3\pi $, $\pi ^+\pi ^-\ga $ and $\pi \ell \nu_{\ell }$ intermediate states into account in the Bell-Steinberger relation, Eq.(2.10) with Eq.(2.11), one may, with the help of Eqs.(2.7a,b) and (2.9), express $\re (\e )$ and $\im (\de )$ in terms of the measured quantities: 
\begin{eqnarray}
\re (\e ) & \simeq & \frac{1}{\sqrt{\ga _S^2+4\De ^2}+4\cos \phi _{SW}\sum _{\ell }\ga _L(\pi \ell \nu _{\ell })}\times \no \\ 
 &&  \Big{[} ~\ga _S(\pi ^+\pi ^-)|\eta _{+-}|\cos (\phi _{+-}-\phi _{SW}) \no \\ 
 && ~~+\ga _S(\pi ^0\pi ^0)|\eta _{00}|\cos (\phi _{00}-\phi _{SW}) \no \\ 
 && ~~+\ga _S(\pi ^+\pi ^-\ga )|\eta _{+-\ga }|\cos (\phi _{+-\ga }-\phi _{SW}) \no \\ 
 && ~~+\ga _L(\pi ^+\pi ^-\pi ^0)\{ \re (\eta _{+-0})\cos \phi _{SW}-\im (\eta _{+-0})\sin \phi _{SW}\} \no \\ 
 && ~~+\ga _L(\pi ^0\pi ^0\pi ^0)\{ \re (\eta _{000})\cos \phi _{SW}-\im (\eta _{000})\sin \phi _{SW}\} \no \\ 
 && ~~+\sum _{\ell }\ga _L(\pi \ell \nu _{\ell })\{ (2\re (x'_{\ell })-d^{\ell }_1(t\gg 1/\ga _S))\cos \phi _{SW} \no \\ 
 && \qquad \qquad \qquad \qquad \qquad \qquad \qquad \qquad -2\im (x'_{\ell })\sin \phi _{SW}\} \Big{]}~, 
\end{eqnarray}
\begin{eqnarray}
\im (\de ) & \simeq & \frac{1}{\sqrt{\ga _S^2+4\De ^2}}\times \no \\ 
 &&  \Big{[} ~-\ga _S(\pi ^+\pi ^-)|\eta _{+-}|\sin (\phi _{+-}-\phi _{SW}) \no \\ 
 && ~~-\ga _S(\pi ^0\pi ^0)|\eta _{00}|\sin (\phi _{00}-\phi _{SW}) \no \\ 
 && ~~-\ga _S(\pi ^+\pi ^-\ga )|\eta _{+-\ga }|\sin (\phi _{+-\ga }-\phi _{SW}) \no \\ 
 && ~~+\ga _L(\pi ^+\pi ^-\pi ^0)\{ \re (\eta _{+-0})\sin \phi _{SW}+\im (\eta _{+-0})\cos \phi _{SW}\} \no \\ 
 && ~~+\ga _L(\pi ^0\pi ^0\pi ^0)\{ \re (\eta _{000})\sin \phi _{SW}+\im (\eta _{000})\cos \phi _{SW}\} \no \\ 
 && ~~+2\sum _{\ell }\ga _L(\pi \ell \nu _{\ell })\{ \re (y_{\ell })\sin \phi _{SW}+\im (x'_{\ell })\cos \phi _{SW}\} \Big{]}~.
\end{eqnarray}
In deriving these equations, use has been made of the fact $\ga _S \gg \ga _L$ . 


\section{Numerical results }

~~~~As the first set of input data, we use the PDG-1996 data[8] as far as available (except those on $\eta _{+-0}$ and $\eta _{000}$) and supplement them by Chell-Olsson's value[15] on $\de _2-\de _0$ and CPLEAR's results[6] on $\eta _{+-0}$, $\eta _{000}$, $d_1^{\ell }(t\gg 1/\ga _S)$ and $d_2^{\ell }(t\gg 1/\ga _S)$. As the second set, we use the CPLEAR data[6] as far as available and supplement them by Gasser-Meissner's value[16] on $\de _2-\de _0$ and the PDG-1996 data[8] for the rest. All the relevant data are recapitulated in Table 1.

\begin{table}[htbp]
	\begin{center}
		\begin{tabular}{c|c|c|c|c}
		\hline \hline 
		&Quantity &PDG-1996 &CPLEAR &Unit \\ \hline 
		&$1/\ga _S$ &$0.8927\pm 0.0009$ & &$10^{-10}s$ \\ 
		&$1/\ga _L$ &$5.17\pm 0.04$ &&$10^{-8}s$ \\ 
		&$-\De $ &$0.5304\pm 0.0014$ &$0.5292\pm 0.0019$ &$10^{10}s^{-1}$ \\ \hline 
		$2\pi $& $\ga _S(\pi ^+\pi ^-)/\ga _S$ & $68.61\pm 0.28$ && $\% $ \\ 
		&$\ga _S(\pi ^0\pi ^0)/\ga _S$ & $31.39\pm 0.28$ & &\% \\ 
		&$|\eta _{+-}|$ &$2.285\pm 0.019$ &$2.316\pm 0.039$ &$10^{-3}$ \\ 
		&$\phi _{+-}$ &$43.7\pm 0.6$ &$43.5\pm 0.8$ &$ ^\circ $ \\ 
		&$|\eta _{00}|$ &$2.275\pm 0.019$ &$2.49\pm 0.46$ &$10^{-3}$ \\ 
		&$\phi _{00}$ &$43.5\pm 1.0$ &$51.7\pm 7.3$ &$ ^\circ $ \\ \hline 
		$3\pi $ &$\ga _L(\pi ^+\pi ^-\pi^0)/\ga _L$ &$12.56\pm 0.20$ &&$\% $ \\ 
		&$\ga _L(\pi ^0\pi ^0\pi^0)/\ga _L$ &$21.12\pm 0.27$ && $\% $ \\ 
		&$\re (\eta _{+-0})$ & &$-0.004\pm 0.008$& \\ 
		&$\im (\eta _{+-0})$ &&$-0.003\pm 0.010$ & \\ 
		&$\re (\eta _{000})$ &&$0.15\pm 0.30$& \\ 
		&$\im (\eta _{000})$ &&$0.29\pm 0.40$ & \\ \hline 
		$\pi \ell \nu $ &$\sum_{\ell }\ga _L(\pi \ell \nu )/\ga _L$ &$65.96\pm 0.30$ &&$\% $ \\ 
		&$\re (x_{\ell })$ &$0.006\pm 0.018$ &$0.0085\pm 0.0102$ & \\ 
		&$\im (x_{\ell })$ &$-0.003\pm 0.026$ &$0.0005\pm 0.0025$& \\ 
		&$d^{\ell }_L$ &$3.27\pm 0.12$ &&$10^{-3}$ \\ 
		&$d^{\ell }_1(t\gg 1/\ga _S)/4$ &&$-1.57\pm 0.70$ &$10^{-3}$ \\ 
		&$d^{\ell }_2(t\gg 1/\ga _S)/4$ &&$-0.07\pm 0.70$ &$10^{-3}$ \\ \hline 
		$\pi ^+\pi ^-\ga $ &$\ga _S(\pi ^+\pi ^-\ga )/\ga _S$ &$0.178\pm 0.050$ &&$\% $ \\ 
		&$|\eta _{+-\ga }|$ &$2.35\pm 0.07$ &&$10^{-3}$ \\ 
		&$\phi _{+-\ga }$ &$44\pm 4$ &&$ ^\circ $ \\ \hline \hline 
		&&Chell-Olsson & Gasser-Mei\ss ner & \\ \hline 
		$2\pi $ &$\de _2-\de _0$ &$-42\pm 4$ &$-45\pm 6$ &$ ^\circ $ \\ \hline 
		\end{tabular}
	\end{center}
	\caption{Input data. }
	\label{1}
\end{table}

Our analysis goes as follows. 

First, assuming $\omega $ being real (see Eq.(5.1)), we use Eqs.(3.8) to find $\omega $. Equations (3.6a,b) are then used to estimate $\eta _0$ and $\eta _2-\eta _0$. The value of $\phi _{SW}$ is obtained from Eq.(2.15). The results are shown in Table 2.

\begin{table}[htbp]
	\begin{center}
		\begin{tabular}{c|c|c|c}
		\hline \hline 
		Quantity &PDG-1996 &CPLEAR &Unit \\ \hline 
		$\omega $ &$2.814\pm 0.357$ &$2.961\pm 0.455$ &$10^{-2}$ \\ 
		$\phi _{SW}$ &$43.44\pm 0.08$ &$43.38\pm 0.11$ &$ ^\circ $ \\ 
		 &&& \\
		$\re (\eta _0)$ &$1.652\pm 0.018$ &$1.632\pm 0.124$ &$10^{-3}$ \\ 
		$\im (\eta _0)$ &$1.575\pm 0.018$ &$1.709\pm 0.136$ &$10^{-3}$ \\ 
		&&& \\ 
		$\re (\eta _2-\eta _0)$ &$-0.121\pm 0.765$ &$5.55\pm 5.13$ &$10^{-3}$ \\ 
		$\im (\eta _2-\eta _0)$ &$0.173\pm 0.446$ &$-2.387\pm 7.238$ &$10^{-3}$ \\ \hline 
		\end{tabular}
	\end{center}
	\caption{Intermediate results. }
	\label{2}
\end{table}

Next, we use Eqs.(5.3a) and (5.4) to estimate $\re (\e )$ and $\re (y_{\ell })$. For this purpose, in view of lack of accurate independent data on $x_{\ell +}$ and $x_{\ell -}$[6], we shall unwillingly neglect possible violation of $CPT$ symmetry in the $\De S \neq \De Q$ amplitudes and assume 
\begin{equation}
x'_{\ell }=0~.
\end{equation}
By combining with Eqs.(5.2a), (5.3b) and (5.5), one may further estimate $\re (\de )$, $\im (\de )$, $\re (y_0)$ and $\im (\e )+z_0$. Equation (5.2b) gives $\re (y_2-y_0)$ and $z_2-z_0$ directly. All the results are compiled in Table 3, where the values of $\de _{\|}$ and $\de _{\perp }$ are also shown.

\begin{table}[htbp]
	\begin{center}
		\begin{tabular}{c|c|c|c|c}
		\hline \hline 
		&&PDG-1996 &CPLEAR &Remark \\ \hline 
		$CP/T$ &$\re (\e )$ &$1.639\pm 0.098$ &$1.695\pm 0.141$ &$\times 10^{-3}$ \\ 
		Violating &$\im (\e )+z_0$ &$1.647\pm 0.102$ &$1.707\pm 0.182$ &$\times 10^{-3}$ \\ 
		&$z_2-z_0$ &$0.173\pm 0.446$ &$-2.38\pm 7.24$ &$\times 10^{-3}$ \\ 
		&$\im (x_{\ell })$ &$-0.3\pm 2.6$ &$0.05\pm 0.25$ &$\times 10^{-2}$; input \\ \hline 
		$CP/CPT$ &$\re (\de )$ &$0.01\pm 9.94$ &$-0.55\pm 9.99$ &$\times 10^{-4}$ \\ 
		Violating &$\im (\de )$ &$0.73\pm 1.01$ &$-0.02\pm 1.21$ &$\times 10^{-4}$ \\ 
		&$\de _{\| }$ &$0.51\pm 7.25$ &$-0.42\pm 7.31$ &$\times 10^{-4}$ \\ 
		&$\de _{\perp }$ &$0.52\pm 6.87$ &$0.36\pm 6.92$ &$\times 10^{-4}$ \\ 
		&$\re (y_0)$ &$0.01\pm 1.00$ &$-0.12\pm 1.02$ &$\times 10^{-3}$ \\ 
		&$\re (y_2-y_0)$ &$-0.121\pm 0.765$ &$5.55\pm 5.13$ &$\times 10^{-3}$ \\ 
		&$\re (y_{\ell })$ &$-0.13\pm 1.41$ &$-0.25\pm 1.42$ &$\times 10^{-3}$ \\ 
		&$\re (x'_{\ell })$ &$0$ &$0$ & assumed \\ \hline 
		$T/CPT$ Violating &$\im (x'_{\ell })$ &$0$ &$0$ & assumed \\ \hline 
		\end{tabular}
	\end{center}
	\caption{Experimental constraints to the symmetry-violating parameters. }
	\label{3}
\end{table}


\section{Comparison with other analyses }

~~~~If one treats $|\omega '|$ as a small quantity, one gets from Eq.(3.6a,b) 
{
\setcounter{enumi}{\value{equation}}
\addtocounter{enumi}{1}
\setcounter{equation}{0}
\renewcommand{\theequation}{\thesection.\theenumi\alph{equation}}
\begin{equation}
\eta _{+-}\simeq \eta _0+\e '~, 
\end{equation}
\begin{equation}
\eta _{00}\simeq \eta _0-2\e ' ~, 
\end{equation}
\setcounter{equation}{\value{enumi}}}%
where 
\begin{equation}
\e '= \frac{1}{\sqrt{2}}\left( \eta _2-\eta _0 \right) \omega \exp (i(\de _2-\de _0))~.
\end{equation}
If one further assumes $CPT$ symmetry, one has, from Eqs.(5.2a,b), 
{
\setcounter{enumi}{\value{equation}}
\addtocounter{enumi}{1}
\setcounter{equation}{0}
\renewcommand{\theequation}{\thesection.\theenumi\alph{equation}}
\begin{equation}
\eta _0=\e +iz_0~,
\end{equation}
\begin{equation}
\e '=\frac{1}{\sqrt{2}}i(z_2-z_0)\omega \exp (i(\de _2-\de _0))~.
\end{equation}
\setcounter{equation}{\value{enumi}}}%

In a note [17] cited in [8], assuming $CPT$ symmetry, the mixing parameter is parametrized as 
{
\setcounter{enumi}{\value{equation}}
\addtocounter{enumi}{1}
\setcounter{equation}{0}
\renewcommand{\theequation}{\thesection.\theenumi\alph{equation}}
\begin{equation}
\frac{q}{p}=\frac{1-\et }{1+\et }~,
\end{equation}
and the $2\pi $ decay amplitude is simply denoted as 
\begin{equation}
\bra (2\pi )_I|\Hw |\ko \ket =A_I~.
\end{equation}
\setcounter{equation}{\value{enumi}}}%
Comparing with Eqs.(2.6), (3.1), (3.2) and (5.1) and with the help of Eqs.(2.13b) and (2.15), one may verify[2] that 
{
\setcounter{enumi}{\value{equation}}
\addtocounter{enumi}{1}
\setcounter{equation}{0}
\renewcommand{\theequation}{\thesection.\theenumi\alph{equation}}
\begin{eqnarray}
 \et &=& \frac{\e \cos \al _K/2-i\sin \al _K/2}{\cos \al _K/2-i\e \sin \al _K/2} \no \\ 
 &=& \frac{2\re (\e )-i \{ (1-|\e |^2)\sin \al _K-2\im (\e )\cos \al _K\} }{1+|\e |^2+(1-|\e |^2)\cos \al _K+2\im (\e )\sin \al _K}~,
\end{eqnarray}
\begin{equation}
\frac{\re (A_2)}{\re (A_0)}=\frac{\omega (\cos \al _K/2-z_2\sin \al _K/2)}{\cos \al _K/2-z_0\sin \al _K/2}~,
\end{equation}
\begin{equation}
\frac{\im (A_I)}{\re (A_I)}=\frac{z_I\cos \al _K/2+\sin \al _K/2}{\cos \al _K/2-z_I\sin \al _K/2}~.
\end{equation}
\setcounter{equation}{\value{enumi}}}%
$\im (\et )$ and $\im (A_I)$ are {\it not} small in general. If, and only if, one restricts himself to the case in which $\al _K$ is allowed to be treated as small as our $\e $ and $z_I$, one has 
{
\setcounter{enumi}{\value{equation}}
\addtocounter{enumi}{1}
\setcounter{equation}{0}
\renewcommand{\theequation}{\thesection.\theenumi\alph{equation}}
\begin{equation}
\et \simeq \e -i\al _K/2 ~,
\end{equation}
\begin{equation}
\frac{\re (A_2)}{\re (A_0)}\simeq \omega ~,
\end{equation}
\begin{equation}
\frac{\im (A_I)}{\re (A_I)}\simeq z_I+\al _K/2~,
\end{equation}
\setcounter{equation}{\value{enumi}}}%
and $\eta _0$ and $\e '$ may be expressed as 
{
\setcounter{enumi}{\value{equation}}
\addtocounter{enumi}{1}
\setcounter{equation}{0}
\renewcommand{\theequation}{\thesection.\theenumi\alph{equation}}
\begin{equation}
\eta _0\simeq \et +i\frac{\im (A_0)}{\re (A_0)}~,
\end{equation}
\begin{equation}
\e '\simeq \frac{i}{\sqrt{2}}\frac{\re (A_2)}{\re (A_0)}\left[ \frac{\im (A_2)}{\re (A_2)}-\frac{\im (A_0)}{\re (A_0)} \right] \exp (i(\de _2-\de _0))~.
\end{equation}
\setcounter{equation}{\value{enumi}}}%
The parameters $\et $ and $A_I$ are not invariant under the rephasing of the $|\ko \ket $ and $|\kob \ket $ states, Eq.(1.1), and it is possible by a choice of phase convention to set $\al _K$ or $\im (A_0)$ or $\im (A_2)$ or $\im (\et )$ to $0$. If one adopts the choice $\im (A_0)=0$ (i.e. the Wu-Yang phase convention[18]), one will have 
{
\setcounter{enumi}{\value{equation}}
\addtocounter{enumi}{1}
\setcounter{equation}{0}
\renewcommand{\theequation}{\thesection.\theenumi\alph{equation}}
\begin{equation}
\eta _0=\et ~,
\end{equation}
\begin{equation}
\e '=\frac{i}{\sqrt{2}}\frac{\im (A_2)}{\re (A_0)}\exp (i(\de _2-\de _0))~.
\end{equation}
\setcounter{equation}{\value{enumi}}}%
Note however that the choice $\im (A_0)=0$ is, as seen from Eq.(7.5c) or (7.6c), equivalent to the choice $z_0=-\tan \al _K/2$ or $\al _K\simeq -2z_0$ and hence Eqs.(7.6a,c) give 
{
\setcounter{enumi}{\value{equation}}
\addtocounter{enumi}{1}
\setcounter{equation}{0}
\renewcommand{\theequation}{\thesection.\theenumi\alph{equation}}
\begin{equation}
\et \simeq \e +iz_0 ~,
\end{equation}
\begin{equation}
\frac{\im (A_2)}{\re (A_2)}\simeq z_2-z_0~.
\end{equation}
\setcounter{equation}{\value{enumi}}}%
Inserting Eqs.(7.6b) and (7.9a,b) into Eqs.(7.8a,b), one goes back to Eqs.(7.3a,b).\footnote{We expect that Eqs.(7.1a,b) with $\eta _0$ and $\e '$ given either by Eqs.(7.7a,b) or Eqs.(7.8a,b) should coincide exactly with Eqs.(4a,b) and (5) derived in [17]. It seems to us that a term propotional to $\im (A_0)$ is missing in their Eqs.(4a,b) or has to be omitted from their Eq.(5). See also Eqs.(13.4a,b,c) of [19] in this respect.}

Such an approximate relation as 
\begin{equation}
\frac{\e '}{\et }\simeq \re \left( \frac{\e '}{\et } \right) \simeq 
\frac{1}{3}\left( 1- \left| \frac{\eta _{00}}{\eta _{+-}} \right| \right) ~,
\end{equation}
is used to estimate $\e '/\et $ (or $\e '/\eta _0$ in our notation) in [8, 17]. On the other hand, from Eqs.(7.1a,b) and the first set of the data in Table 1, 2 and 3, we find 
\begin{equation}
\left. 
	\begin{array}{c}
	\re \left( \frac{\e '}{\eta _0} \right) = (1.5\pm 5.5)\times 10^{-3}~, \\ \\ 
	\im \left( \frac{\e '}{\eta _0} \right) =(1.2\pm 5.5)\times 10^{-3}~.
	\end{array}
\right. 
\end{equation}
It appears therefore that the second (first) near equality in Eq.(7.10) is (may not be) justifiable.\footnote{It is argued that the first near equality in Eq.(7.10) follows, since 
\begin{equation}
\mbox{The phase of } ~\et ~\mbox{ in the Wu-Yang phase convention}\simeq \phi _{SW}~,
\end{equation}
and since $\phi _{SW}$ is accidentally close to the phase of $\e '$, $\de _2-\de _0+\pi /2$. Note however that Eq.(7.12) holds only when direct $CP$ violation (i.e. $CP$ violation in decay amplitudes and decay matrix) is negligible (see Eqs.(2.13b), (2.15) and (7.9a)) and that $\e '$ {\it is} a quantity related to direct $CP$ violation. 
} 

In [8], 
{
\setcounter{enumi}{\value{equation}}
\addtocounter{enumi}{1}
\setcounter{equation}{0}
\renewcommand{\theequation}{\thesection.\theenumi\alph{equation}}
\begin{equation}
\zeta =\frac{m_{\ko }-m_{\kob }}{m_{\ko }}~,
\end{equation}
is quoted as a typical quantity which signals $CPT$ violation. $\zeta $ is related to our parameters defined in Eq.(2.14b) as 
\begin{equation}
\zeta =\frac{\de _{\perp }\sqrt{\ga _S^2+4\De ^2}}{m_{\ko }}~,
\end{equation}
\setcounter{equation}{\value{enumi}}}%
and it is true that $\zeta \neq 0$ would imply violation of both $CP$ and $CPT$ symmetries. It seems however that $\de _{\perp }$ itself (rather than $\zeta $) is better to be regarded as a parameter which characterizes symmetry violation[13, 20]. In [21] cited in [8], a couple of assumptions and approximations are made to relate $\de _{\perp }$ directly to the measured quantities such as $\eta _{+-}$, $\eta _{00}$ and $\phi _{SW}$. Among the asumptions is direct $CPT$ violation being negligible, which would however at the same time lead to $\de _{\| }=0$. It appears that, in view of our numerical results shown in Table 3, such an assumption may not be justifiable. \footnote{A similar remark was also raised in [22].}


\section{Concluding remarks }

~~~~We have introduced a set of parameters to describe possible violation of $CP$, $T$ and $CPT$ symmetries and of the $\De S=\De Q$ rule in the $\ko $-$\kob $ system in a well-defined way and attempted to derive constraints to these parameters from the presently available experimental data in a way as phenomenological and comprehensive as possible.

From our numerical results shown in Table 3, it is seen that, in contrast to $\re (\e )$ and $\im (\e )+z_0$, which are definitely non-vanishing and are of the order of $10^{-3}$, all the other symmetry-violating parameters are consistent with being vanishing and are at most of the order of $10^{-3}$. This implies, on the one hand, that all the present observations are consistent with no $CPT$ violation  and no direct $CP$ and $T$ violations, and, on the other hand, that $CPT$ violation and direct $CP$ and $T$ violations up to a level comparable to that of indirect $CP$ and $T$ violations are at present not excluded. It is therefore not advisable to neglect direct symmetry violation in phenomenological analyses. 

We have to admit that our analysis is not totally free from theoretical prejudices and is subject heavily to experimental uncertainties, among which we mention: 

(1) We have unwillingly accepted Eq.(6.1). In this respect, we would like to stress that measurements on various leptonic asymmetries without this or that theoretical inputs are highly desirable and that $\im (x'_{\ell })$ is the only parameter which characterizes $T$ and $CPT$ violation but has nothing to do with $CP$ violation. 

(2) We have treated the $3\pi $ ($\pi ^+\pi ^-\ga $) state as purely $CP$-odd ($CP$-even) and taken these states into account when using the Bell-Steinberger relation to estimate $\re (\e )$ and $\im (\de )$. As a result, our final numerical results are subject to uncertainties which come largely from experimental errors on $\eta _{+-0}$, $\eta _{000}$ and $\eta _{+-\ga }$. It is hoped that, in the near future, more abundant and accurate data on these and other relevant quantities will become available and enable one to identify and/or constrain $CP$, $T$ and/or $CPT$ violations in a more precise way.

(3) The Bell-Steinberger relation has played a very important role in our analysis. It is to be noted in this respect that fully time-dependent measurements on leptonic asymmetries of various types will allow one to identify or constrain $\re (\e )$ and $\im (\de )$ and thereby test this relation itself[12, 14].

Finally, as mentioned earlier, our parametrization is not fully rephasing invariant. A more thorough discussion of phase ambiguities associated with final state as well as the $\ko $ and $\kob $ states will be given elsewhere[14]. 


\section*{Acknowlegements }

~~~~One of the present authors (S. Y. T.) would like to express his thanks to T. Morozumi for giving an opportunity to report the present work at the Workshop on Fermion Masses and $CP$ violation held in Hiroshima on March 5-6, 1998. He is also grateful to P. Pavlopoulos for bringing Ref.[7] to his attention and to Z.Z. Xing for useful communications. 


\end{document}